\documentclass[twocolumn,superscriptaddress]{revtex4-1}

\usepackage{amsfonts,amsmath,amssymb}
\usepackage{graphicx}
\usepackage[usenames,dvipsnames]{xcolor}

\usepackage{lettrine}
\renewcommand{\figurename}{\textbf{Figure}}
\def\fnum@figure{\figurename\nobreakspace\textbf{\thefigure}}

\newcommand{\as}[1]{\hat{a}_\mathrm{s}}
\newcommand{\As}[1]{\hat{a}_\mathrm{s}^\dag}

\newcommand{\blank}{\vspace{3mm}\noindent}

\begin{document}

\title{Monitoring Microtubule Mechanical Vibrations via Optomechanical Coupling}
\author{Sh. Barzanjeh}
\affiliation{Institute of Science and Technology~(IST) Austria, 3400 Klosterneuburg, Austria}
\author{V. Salari}
\affiliation{School of Physics, Institute for Research in Fundamental Sciences (IPM), Tehran 19395-5531, Iran}
\author{J. A. Tuszynski}
\affiliation{Department of Oncology, University of Alberta, Cross Cancer Institute, Edmonton T6G 1Z2, Alberta, Canada}
\affiliation{Department of Physics, University of Alberta, Edmonton,  AB T6G 2E1, Canada}
\author{M. Cifra}
\affiliation{Institute of Photonics and Electronics, The Czech Academy of Sciences, Chabersk\'{a} 57, 182 00 Prague, Czech Republic}
\author{C. Simon}
\affiliation{Department of Physics and Astronomy, University of Calgary, Calgary T2N 1N4, Alberta, Canada}
\affiliation{Institute for Quantum Science and Technology, University of Calgary, Calgary T2N 1N4, Alberta, Canada}

\begin{abstract}
The possible disruption of a microtubule during mitosis can control the duplication of a cancer cell. Cancer detection and treatment may be possible based on the detection and control of microtubule mechanical oscillations in cells through external fields (e.g. electromagnetic or ultrasound). However, little is known about the dynamic (high-frequency) mechanical properties of microtubules. Here we propose to control the vibrations of a doubly clamped microtubule by tip electrodes and to detect its motion via the optomechanical coupling between the vibrational modes of the microtubule and an optical cavity. In the presence of a red-detuned strong pump laser, this coupling leads to optomechanical induced transparency of an optical probe field, which can be detected with state-of-the art technology. The center frequency and linewidth of the transparency peak give the resonance frequency and damping rate of the microtubule respectively, while the height of the peak reveals information about the microtubule-cavity field coupling. Our method should yield new knowledge about the physical properties of microtubules, which will enhance our capability to design physical cancer treatment protocols as alternatives to chemotherapeutic drugs. 
\clearpage\end{abstract}

\maketitle

%


\lettrine{M}icrotubules~(MTs) play essential roles in many fundamental physiological processes in all eukaryotic cells. The most prominent role of MTs is to provide the mechanical force required for chromosome separation during mitosis. MTs are self-assembled from tubulin hetero-dimers as hollow cylinders having 25 nm external and 15 nm internal diameter. The length of MTs can vary from tens of nanometers to hundreds of microns~\cite{Civalek}. 
Due to the central role of MTs in mitosis, a disruption of MT function during mitosis can arrest cell division, so any physical or chemical mechanism that affects MT assembly can potentially be useful as a modality in the treatment of cancer, for example high-frequency ultrasound or AC electric fields~\cite{Samarbakhsh, Samarbakhsha, Kirson}. The mechanical stiffness of individual MTs controls the mechanical properties of the cytoskeletal network, and it is modified for different functions in the cell~\cite{Hawkins}. Theoretical and experimental analysis of MT structure reveals their capability for mechanical oscillations over a wide frequency range from acoustic to GHz frequencies~\cite{Sahu, Sirenko, Wang, Qian, Tuszynski}. 

Highly dynamic mitotic-spindle MTs are among the most successful targets for anticancer therapy~\cite{Dumontet, Jordan}. Now the question is: can we induce disruption of MTs in cancer cells via external physical fields such as intense pulsed ultrasound or electric field, instead of chemical drugs, to cause the least possible damage to the healthy cells? Here, we suggest that having solid knowledge about the mechanical vibrations of MTs is important in order to reach this goal. While there are multiple theoretical and experimental studies on MT static mechanical properties \cite{Pablo,Felgner,Gittes,Kikumoto,Kis,Kurachi,Mickey,Needleman,Pampaloni,Venier}, information on dynamic high-frequency properties of MT comes almost exclusively from theory \cite{Beni, Daneshmand,Demir,Deriu,Ghavanloo,Ghorbanpour}, with little experimental data available for comparison \cite{Hameroff86,Pizzi}. Having knowledge of MT dynamic properties would also shed light on hypotheses of MT based electrodynamic cellular signaling \cite{Priel} and information processing \cite{Hameroff14}, and on controversial results concerning peculiar high-frequency electronic resonances of MTs \cite{Sahu,Sahua,Sahub}. Regardless of whether new experimental results on microtubule dynamic properties support or reject these hypotheses, such results would significantly enhance our capabilities to rationally design physical methods to influence microtubule-based cellular functions such as mitosis, potentially leading to new cancer treatment protocols. However, no technique has been demonstrated to be capable of probing high frequency (MHz - GHz) mechanical properties of individual MTs.

In this paper we propose a new technique for the analysis of MT high frequency dynamics at room temperature by using the optomechanical coupling in an optical cavity, similar to what has been done for nanomechanical resonators~\cite{Kipp3}. Cavity-based optomechanical systems display a parametric coupling between the
displacement $X$ of a mechanical vibration mode and the energy stored inside a radiation mode. Nano-electromechanical and optomechanical resonators~\cite{Aspelmeyer, Blencowe, Favero, barzanjehnano} have different applications including the sensitive detection of physical quantities such as spin~\cite{Rugar, Budakian}, monitoring biological samples~\cite{Huber}, testing quantum mechanical behavior at the mesoscopic to macroscopic level~\cite{O'Connell,Teufel,Poot,Barzanjeh3,Marshall,Pepper,Ghobadi}, and frequency conversion~\cite{Barzanjeh1, Barzanjeh2, Andrew, barzanjeh4}. A driving laser applied to the red mechanical sideband of an optical cavity can be used to slow down and even stop light signals due to optomechanically induced transparency~\cite{Weis, Safavi, Agarwal, Ma}. On the other
hand, driving the optomechanical cavity on the blue sideband can lead to phonon lasing~\cite{Rokhsari,Herrmann} and probe amplification~\cite{Safavi,Massel, Metelmann}. 

In this work, first we present a model describing the vibrations of a MT based on \textit{Euler-Bernoulli} beam theory. We consider a doubly clamped MT and find both the Lagrangian and the Hamiltonian of the MT. Furthermore, the dielectric properties of the MT afford a new opportunity to control and modulate the MT vibrations with an electrostatic gradient force, which originates from an inhomogeneous external electric field. We show that both vibrational and equilibrium position of the MT can be modulated by positioning tip electrodes close to the center of a doubly clamped MT. Finally, we propose a new method to read out the information associated with the MT vibrations by coupling the MT to an optical cavity. The optomechanical coupling between the optical cavity and the vibrational mode of the microtubule in the presence of an external driving force induces a strong optical sideband field due to the anti-Stokes scattering of light. When the external-force frequency~($ \omega $) is close to the characteristic MT mechanical frequency~($ \Omega_m$), the driving force coherently enhances the oscillation of the MT, inducing a strong optical sideband field and leading to optomechanically induced transparency in the transmission of the probe field. 

\begin{figure}[h]
\centering
\includegraphics[width=0.9\columnwidth]{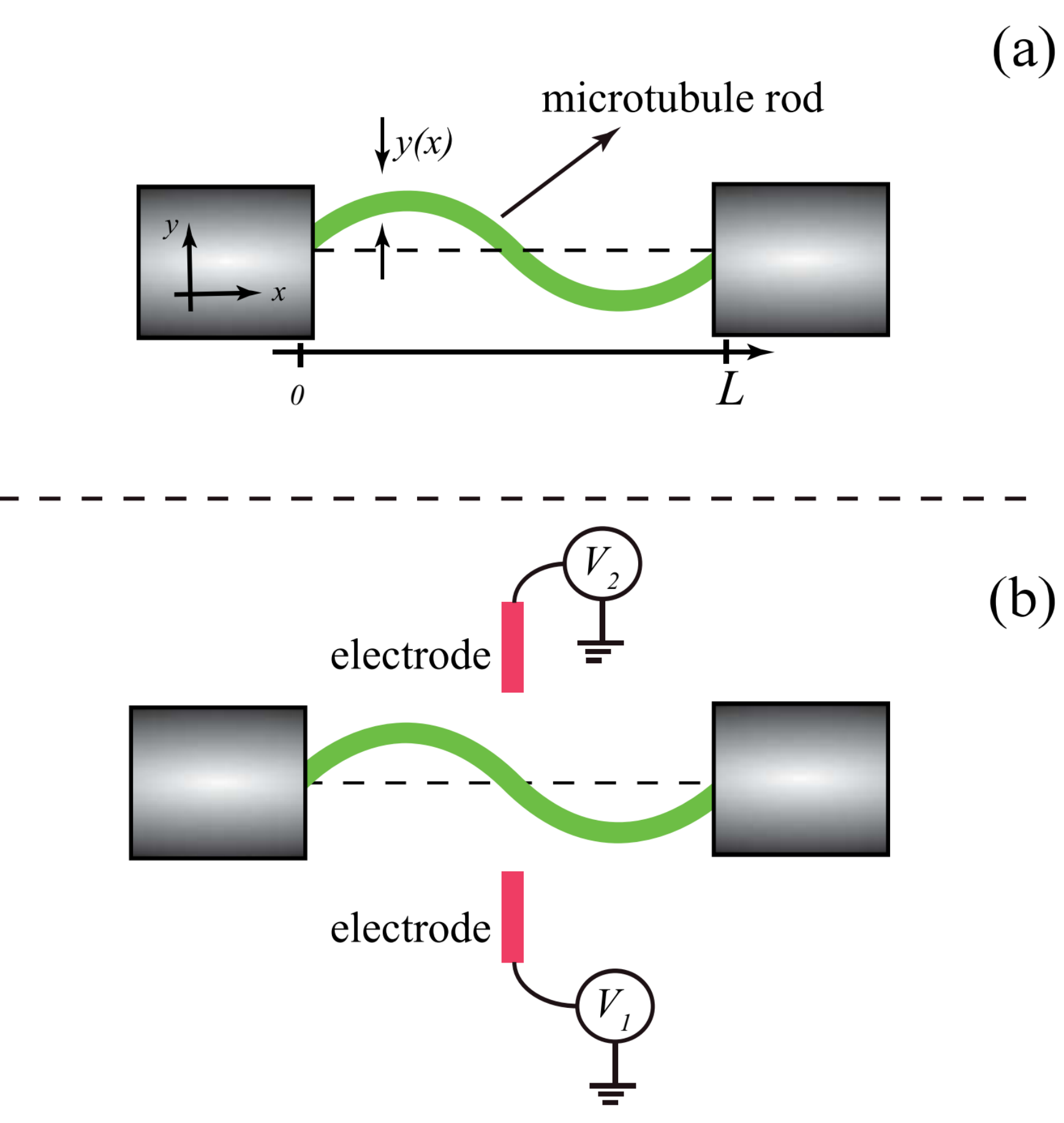}
\caption{{\bf Doubly-clamped microtubule.} (a)~Schematic description of a doubly clamped microtubule with length $ L $. (b)~Controlling the microtubule vibrations by applying electric fields. Here, we assume the tip electrodes are placed close to the center of the MT.}
 \label{fig1}
\end{figure}

\blank
\textbf{Results}\\
\textbf{Elasticity and vibrational modes of the Microtubule.}~~The vibrations of the MT can be fully charactrized via the \textit{Euler-Bernoulli} beam theory. In this paper, we consider a doubly clamped MT with a constant linear mass density $ \mu $ and length~$ L $  in which the radius of circular cross-section $ R $ is much smaller than the length viz,. $ L\gg R $~[See Fig.~\ref{fig1}(a)]. The MT is assumed to be homogeneous along the longitudinal axis $ x\in [0,L] $ in which the planar deflection in the transverse direction is described by $ y(x) $. Based on the \textit{Euler-Bernoulli} beam theory, the Lagrangian of the MT can be written as follows~\cite{cleland, rips, hartman}
\begin{equation}\label{lag}
\mathcal{L}[y(x)]=\frac{\mu}{2}\int dx \frac{d y(x)}{dt}-\frac{1}{2}\int Y\,A  \Big[\sigma\frac{d^2 y(x)}{dx^2}\Big]^2,
\end{equation}
where $ A $ is the cross-sectional area, $ Y $ is the Young modulus, and $ \sigma $ is the ratio between the bending and compressional rigidity of the MT. For a cylindrical shell like a MT with radius $ R $ one finds $\sigma\simeq R/\sqrt{2} $. We note that in the Lagrangian (\ref{lag}), the boundary conditions are obtained by assuming the two-sided clamped MT where the end points at $ x=0 $ and $ x=L $ are fixed, means that $ y(0)=y(L)=0 $ and $y'(0)=y'(L)=0 $. 

The Hamiltonian associated with Lagrangian~(\ref{lag}) describing MT vibrations is given by~(\textbf{See} \textbf{Methods})
\begin{equation}\label{hamiltonianH0}
H_{MT}=\sum_n \Big(\frac{P_n^2}{2m_n}+\frac{1}{2}m_n\omega_n^2X_n^2\Big),
\end{equation}
where $ X_n $ and $ P_n=m_n\frac{\partial X_n}{\partial t} $ are the deflection and mode momentum for the $ n $th mode, respectively, while $ m_n=\mu \int_0^L \psi^2_n(x)dx$ is the effective mode mass with vibrational frequencies $ \omega_n=\sigma \sqrt{Y\,A/\mu}(\zeta_n/L)^2 $. 

Eq. (\ref{hamiltonianH0}) describes the multimode Hamiltonian for the doubly clamped MT. The Hamiltonian of the fundamental mode $ n=1 $ is given by~$
H_{MT}\vert_{n=1}=\frac{P^2}{2m_1}+\frac{1}{2}m\omega_{1}^2X^2$, 
where $ \omega_1 $ is the MT's fundamental frequency and $ m_1\simeq 0.3965 \mu L $ shows the effective mass of the fundamental mode. In general, MTs are considered as hollow cylinders~\cite{Howard, Boal} having 25 nm external and 15 nm internal diameters and their length can vary from $ 10\,n $m to $ 100\,\mu $m. Their mechanical rigidity is quantified by the Young modulus $Y=(1.2-2.5)\times 10^9 \mathrm{N\,m}^{-2}  $~\cite{Wagner, Tolomeo, Tuszynski} and a linear mass density is $ \mu\simeq 3.4\times 10^{-13} \mathrm{kg\,m}^{-1} $~\cite{Gu}. Therefore, depending on the length, the MT's fundamental frequency ranges from $ 100 $ kHz to $ 100$ GHz with the effective mass varying between $ 10^{-17} $kg to $ 10^{-21} $kg. The zero-point fluctuation amplitude is given by $ x_{zpf}=\sqrt{\frac{\hbar}{2m\Omega_m}}$,
which indicates that the spread of the coordinate in the ground state has a value between $ 0.01 $pm to $ 10 $pm. 

\blank

\textbf{Controlling microtubule vibrations with an external field.}
The dielectric properties of MT's afford a new opportunity to control and modulate MT vibrations with an electrostatic gradient force originated from an inhomogeneous external electric field~\cite{Unt} applied to MT's. Here we consider a configuration shown in Fig. \ref{fig1}(b) in which two tip electrodes are placed near the center of a doubly clamped MT. The electrostatic field generated by these tips produces an additional potential acting on the MT with the following energy per unit length along the MT as
\begin{equation}
U_{el}(x,y)=-\frac{1}{2}(\alpha_{\parallel}E^2_{\parallel}+\alpha_{\perp}E^2_{\perp}),
\end{equation}
where $E_{\parallel}\,(E_{\perp})  $ is the external field component parallel~(perpendicular) to the MT axis and  $\alpha_{\parallel}\,(\alpha_{\perp})  $ shows the associated screened polarizability. By expanding the electrostatic energy $ U_{el}(x,y) $ to the first order in the displacement variable $ y $ i.e., $ U_{el}(x,y)\simeq U_{el}(x,0)+y(\frac{\partial U_{el}}{\partial y}|_{y=0})$ and using the modes defined in Eq.~(\ref{solution}), we obtain the Hamiltonian of the electrostatic force acting on the MT
\begin{equation}\label{elastic}
H_{el}=\int_{0}^{L}U_{el}(x,y)dx\simeq \sum_{n}F_n X_n,
\end{equation}
where $F_n=\int_0^L \psi_n\Big(\frac{\partial U_{el}}{\partial y}|_{y=0}\Big) dx$ is the electrostatic force acting on the MT.
Note that in the Hamiltonian (\ref{elastic}) we have ignored the displacement-independent term $ U_{el}(x,0)  $ which only shifts the energy level of the system since it is irrelevant for the MT dynamics. The Hamiltonian~(\ref{elastic}) describes the electrostatic force leading to a static deflection of the MT which causes a shift in its equilibrium position. This raises a new possibility to control MT vibrations. We could also consider time-dependent electric fields acting on MT. In this case, it is more convenient to discriminate the static and time-dependent force contributions, i.e. $ F_d(t)=\bar{F}_{0}+\delta F_d(t) $ where $ \delta F_d(t)$ shows the time-dependent external drive acting on the MT.  

In the next section, we present a setup to observe MT vibrations by using an optical cavity. We show that placing a MT next to an optical cavity changes the resonance frequency of the cavity; leading to the appearance of an optomechanical coupling between the cavity and MT vibrations. 
\blank

\begin{figure}[h]
\centering
\includegraphics[width=1\columnwidth]{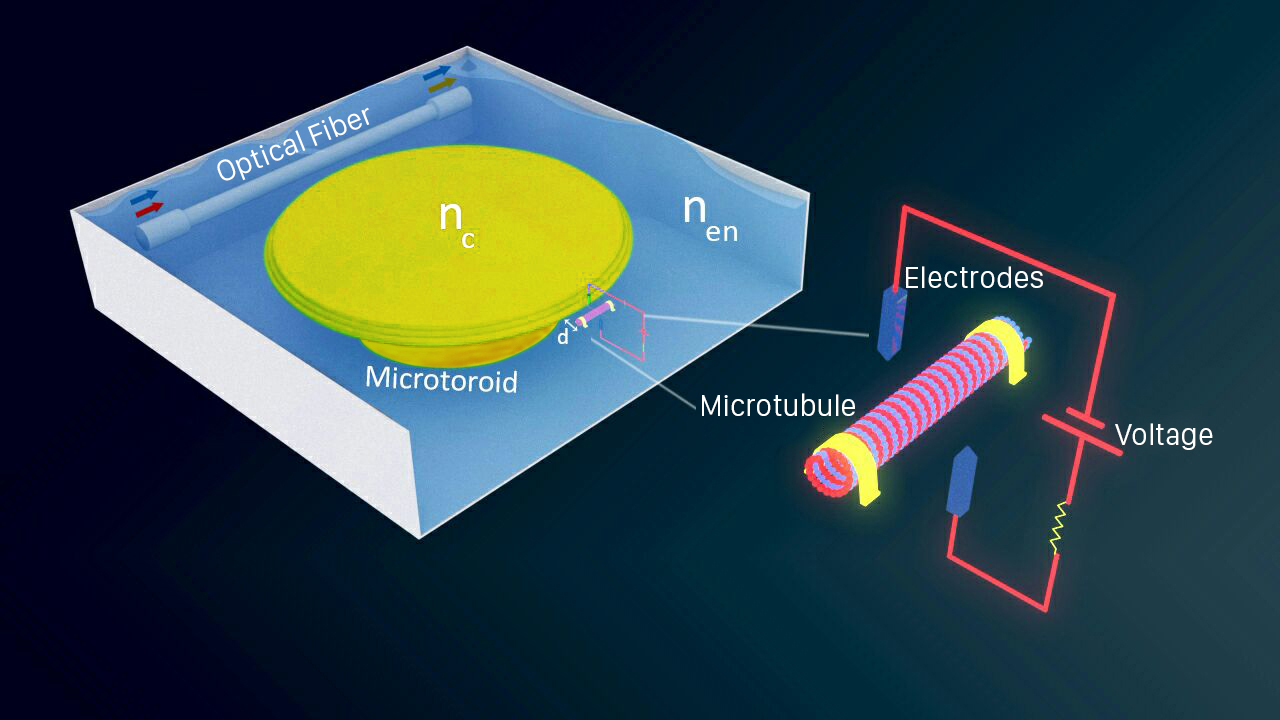}
\caption{\textbf{Sketch of the system.} A doubly-clamped microtubule with mass $ m $ and resonance frequency~$ \Omega_m $ couples to the evanescent field of a whispering gallery mode~(WGM) micro-toroid cavity with refractive index $ n_c $ and resonance frequency $\omega_{c}$. The distance between the microtubule and the cavity rim is $ d $. The MT vibrations can be controlled by an external electric gradient field. One can monitor MT vibrations by observing the output of the optical cavity. The entire system is located inside an aqueous medium with refractive index $ n_{en} $.}
 \label{fig2}
\end{figure}

\textbf{Optomechanical coupling between the microtubule and the optical cavity.
}  The system considered here is shown in Fig.~(\ref{fig2}) in which a whispering gallery mode~(WGM) optical-cavity~\cite{Vahala, Heebner}, with resonance frequency $\omega_{c}$, is placed next to a doubly clamped MT with mass $ m $ and resonance frequency~$ \Omega_m $. The evanescent field of the WGM cavity establishes an optomechanical coupling with the motion of the MT in which the cavity field exerts a radiation pressure force on the MT while the MT displacement from the equilibrium position simultaneously changes the resonance frequency of the cavity. In order to keep the MT stable in the rod form we assume that the whole system is located inside an aqueous medium with refractive index $ n_{en} $. The Hamiltonian of the system describing the MT-cavity coupling is given by
 \begin{equation}\label{Hamtotal}
 \hat H=\hat H_{MT}+\hbar\omega_{c} \hat a^{\dagger}\hat a+\hbar g_{0}\hat X \hat a^{\dagger}\hat a
 \end{equation}
where $ \hat a\,(\hat a^{\dagger}) $ denotes the cavity annihilation (creation) operator
satisfying the commutation relation~$[\hat a,\hat a^{\dagger}]=1  $ while $  \hat P$ and $ \hat X $ are, respectively, the momentum and position operator of the MT. The single mode Hamiltonian $ \hat H_{MT}=\hat P^2/2m+m\Omega_m^2 \hat X^2/2 $  is the quantized version of the MT's Hamiltonian in which we limited our analysis to the fundamental mode. The third term of the Hamiltonian (\ref{Hamtotal}) shows the free energy of the optical cavity while the last term indicates the optomechanical coupling between the cavity field and MT's vibrations in which  the deflection of the MT shifts the resonance frequency of the cavity with a rate given by~[\textbf{see} \textbf{Methods}] 
\begin{equation}\label{couplingg0}
g_0=\frac{\partial \omega_{c}}{\partial X}\simeq\Big(\frac{\omega_c \alpha_{||}k_{\perp}\zeta^2 }{n_c^2 \epsilon_0 V_c}\mathrm{e}^{-2k_{\perp}d}\Big) A_c
\end{equation}
where $ \lambda_c=2\pi c/\omega_c $ and $ n_c $ are the wavelength and refractive index of the microcavity, respectively, and  $k_{\perp}^{-1}=1/\sqrt{n_c^2-n_{en}^2}k $ is the decay length of the evanescent field, with $ k=2\pi/\lambda_c $ being the wavenumber of the field. $n_{en}  $ is the refractive index of the cavity/MT environment where and $ \zeta=\frac{0.42 \lambda_c}{R\sqrt{n_c^2-n_{en}^2}} $ while $ d $ is the distance between the MT center and the cavity rim and $ V_c $ is the mode volumn of the optical mode. The correction term $ A_c\simeq 0.17[k_{\perp}(d+R)]^{-1/2} $ accounts for the misalignment and mispositioning of the MT with respect to the cavity rim~\cite{hartman}. In the derivation of Eq.~(\ref{couplingg0}), we have neglected the contribution of the perpendicular polarizability as we assume the parallel polarizability of the MT provides the main contribution to the optomechanical coupling. We also note that, although we introduced the quantized Hamiltonian for the MT but in this paper we are interested in the mean response of this system; therefore we can analyze the expectation values instead of the operators.  

\blank
\textbf{Microtubule Induced Optical Transparency.}
 In this section, we propose a scheme to measure vibrations of a MT, which is driven by an external time-harmonic driving force $ \delta F_d(t)=f_d \mathrm{cos}(\omega t+\phi_d) $ with frequency $  \omega$, amplitude $ f_d $, and phase $ \phi_d $. The optical cavity is driven by
a strong pump field of frequency $  \omega_l$ and a very weak probe field of frequency $  \omega_p$. We introduce the amplitudes of the pump field
and the probe field inside the cavity $ \mathcal{E}_l=\sqrt{\frac{ P_l}{\hbar \omega_l}} $ and  $ \mathcal{E}_p=\sqrt{\frac{ P_p}{\hbar \omega_p}} $, respectively, where $ P_l $ is the pump power and 
$  P_p$ is the power of the probe field. Similar to Eq.~(\ref{Hamtotal}), the quantized Hamiltonian describing the MT-cavity interaction in the presence of a time-harmonic-driving force (see Fig.~\ref{fig2}) reads
 \begin{eqnarray}\label{QuantizedHam}
 \hat H&=&\frac{\hat{P}^2}{2m}+\frac{1}{2}m\Omega_{M}^2\hat{X}^2+\hbar \omega_{c} \hat{a}^{\dagger}\hat a\\
 &+&\hbar g_{0}\hat X  \hat{a}^{\dagger}\hat a-\big[\bar{F}_0+f_d\, \mathrm{cos}(\omega t+\phi_d)\big] \hat X \nonumber\\
 &+&i\hbar\sqrt{ \kappa_e}\Big[\Big(\mathcal{E}_le^{-i\omega_{l}t}+\mathcal{E}_p e^{-i\omega_{p}t-i\phi_p}\Big) \hat a^{\dagger}-h.c\Big]\nonumber,
\end{eqnarray}
where $ \phi_p $ is the initial phase of the probe field and $ \kappa_{e} $ is the extrinsic damping rates of the cavity.

In a frame rotating at $  \omega_l$ and by considering the quantum and thermal noise of the mechanical oscillator $ \hat F_{th} $ and $ \hat a_{in} $, we obtain
the Heisenberg-Langevin equations of motion associated with Hamiltonian~(\ref{QuantizedHam})
\begin{subequations}\label{Quantizedequation of motion}
\begin{eqnarray}
\dot{\hat{{X}}}&=&\frac{\hat P}{m},\\
\dot{\hat{{P}}}&=&-m\Omega_m^2\hat X-\hbar g_0 \hat{a}^{\dagger}\hat a-\gamma_m \hat P\nonumber\\
&+&\bar{F}_0+f_d\,\mathrm{cos}(\omega t+\phi_d)+\hat{F}_{th},\\
\dot{\hat{a}}&=&\Big(i\Delta_0-\frac{\kappa}{2}\Big)\hat a-ig_0 \hat X \hat a+\sqrt{\kappa_e}(\mathcal{E}_l+\mathcal{E}_p e^{-i\delta t-i\phi_p})+\hat a_{in},\nonumber
\end{eqnarray}
\end{subequations}
where $ \Delta_0=\omega_l-\omega_c $, $ \delta=\omega_p-\omega_l $ and the total cavity
decay rate is $  \kappa=\kappa_{e}+\kappa_i$ in which $ \kappa_{i} $ is the intrinsic damping rate of the cavity. Here, $ \gamma_M$ is the damping rate of the MT and $ \hat{F}_{th} $~(with  $ \langle\hat{F}_{th}\rangle=0 $) denotes the sum of all incoherent external forces (such as random Langevin and/or viscous forces) that are acting on the MT, and obeys $ \langle \hat F_{th}(t)\hat F_{th}(t')\rangle=k_B T m\gamma_M \delta(t-t') $, where~$ k_B $ is the Boltzmann constant and $ T $ is the absolute temperature of the reservoir.

 In the present work, we are interested in the mean response of this system; therefore we can analyze the expectation values instead of the operators. Since $ \langle\hat a_{in}\rangle=0 $ and $\langle \hat F_{th}\rangle=0 $, we find
\begin{subequations}\label{meanEqution}
\begin{eqnarray}
\langle\dot{\hat{a}}\rangle &=&\Big(i\Delta_0-\frac{\kappa}{2}\Big)\langle\hat{a}\rangle -ig_0 \langle\hat{X}\rangle \langle\hat{a}\rangle+\sqrt{\kappa_e}(\mathcal{E}_l+\mathcal{E}_p e^{-i\delta t-i\phi_p}),\nonumber\\
\\
\langle\ddot{\hat{X}}\rangle &+&\gamma_M \langle\hat{\dot{X}}\rangle+\Omega_M^2 \langle\hat{X}\rangle=-\frac{\hbar g_0 |\langle \hat a \rangle|^2}{m}\nonumber\\
&+&\frac{\bar F_0}{m}+\frac{f_d}{m}\mathrm{cos}(\omega t+\phi_d),
\label{meanEqution2}
\end{eqnarray}
\end{subequations}
where $-\hbar g_0 |a|^2 $ is the cavity field radiation pressure acting on the MT.  The above equations are non-linear and one cannot solve them analytically. Thus, we can use perturbation methods to obtain an approximate analytical solution for the case that the probe field is much weaker than the pump field i.e., $ \mathcal{E}_p\ll \mathcal{E}_l $. 

In this paper we are interested in the transmission of the probe field which is the ratio of the probe field returned from the system $ \langle \hat a_{out}\rangle $ divided by the sent probe field~$ \mathcal{E}_p e^{-i(\omega_pt+\phi_p)} $ and it is given by~[\textbf{see Methods}]
\begin{eqnarray}\label{transmision2}
T_p(\omega)&=&1-\frac{\sqrt{\kappa_e}}{\mathcal{E}_pe^{-i\phi_p}}\Big(\alpha_1^-+\alpha_2^-\Big).
\end{eqnarray}
where we assumed the cavity is driven in red-sideband $ \Delta=-\Omega_m $ where $ \Delta=\Delta_0-g_0x_0 $ being the effective optical detuning in which $ x_0 $ is the equilibrium displacement of the MT. In Eq.~(\ref{transmision2}) the real part of $ \frac{\sqrt{\kappa_e}}{\mathcal{E}_pe^{-i\phi_p}}\Big(\alpha_1^-+\alpha_2^-\Big) $ shows the absorptive behavior of the cavity and its imaginary part describes
the dispersive behavior. Note that, the second term of the above equation~($ \sqrt{\kappa_e}\alpha_1^-/\mathcal{E}_pe^{-i\phi_p} $) indicates the effect of the vibrations of the MT on the transmission of the probe field while the third term~($ \sqrt{\kappa_e}\alpha_2^-/\mathcal{E}_pe^{-i\phi_p}$) shows that applying the external force could substantially change the MT vibration and consequently alters the response of the system. In the resolved sideband regime i.e., $ \Omega_m\gg{\kappa} $, Eq.~(\ref{transmision2}) can be simplified further $
T_p(\Omega_m) \simeq 1-2\Big(1-\frac{G_de^{i\phi}}{\sqrt{\kappa_e}\mathcal{E}_p}\Big)
$, where $ G=g_0\sqrt{n_d}X_d $ is the effective MT-cavity field coupling rate, $ \phi=\phi_p-\phi_d $, and $ X_d=\frac{f_d}{m\gamma_m\Omega_m }$ is the total displacement imposed by the time-dependent external force. Here, $ n_d = \frac{4\kappa_e^2\mathcal{E}_l^2}{\kappa^2+4\Delta^2}$ shows the total number of photons inside the cavity. Note that we have chosen $ f_d $ such that $ k_{\perp}X_d \ll 1 $. In principle, Eq.~(\ref{transmision2}) shows that when the external-force frequency is close to $ \Omega_m$, the MT motion driven through the external force induces an optical sideband field due to the anti-Stokes scattering of light, which interferes with the probe field and the anti-Stokes field induced by probe field, leading to the modification of the output field.
 \begin{figure}[h]
\centering
\includegraphics[width=0.9\columnwidth]{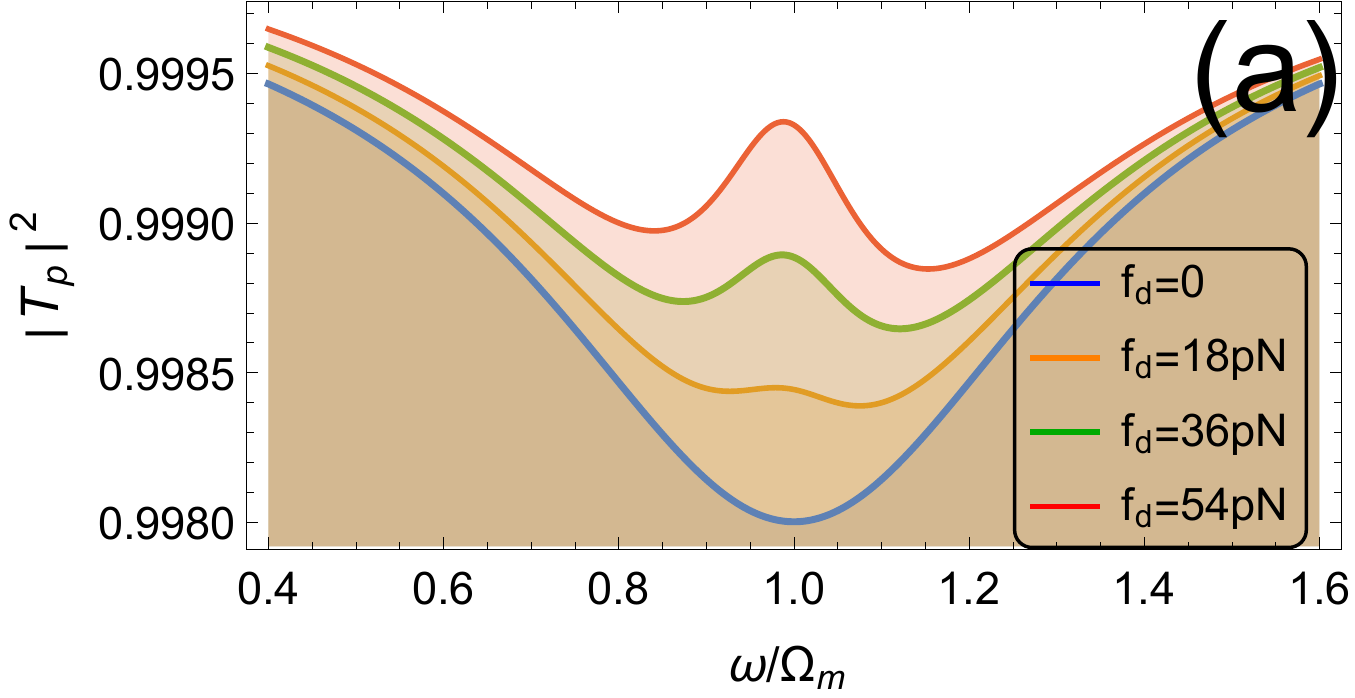}
\includegraphics[width=0.9\columnwidth]{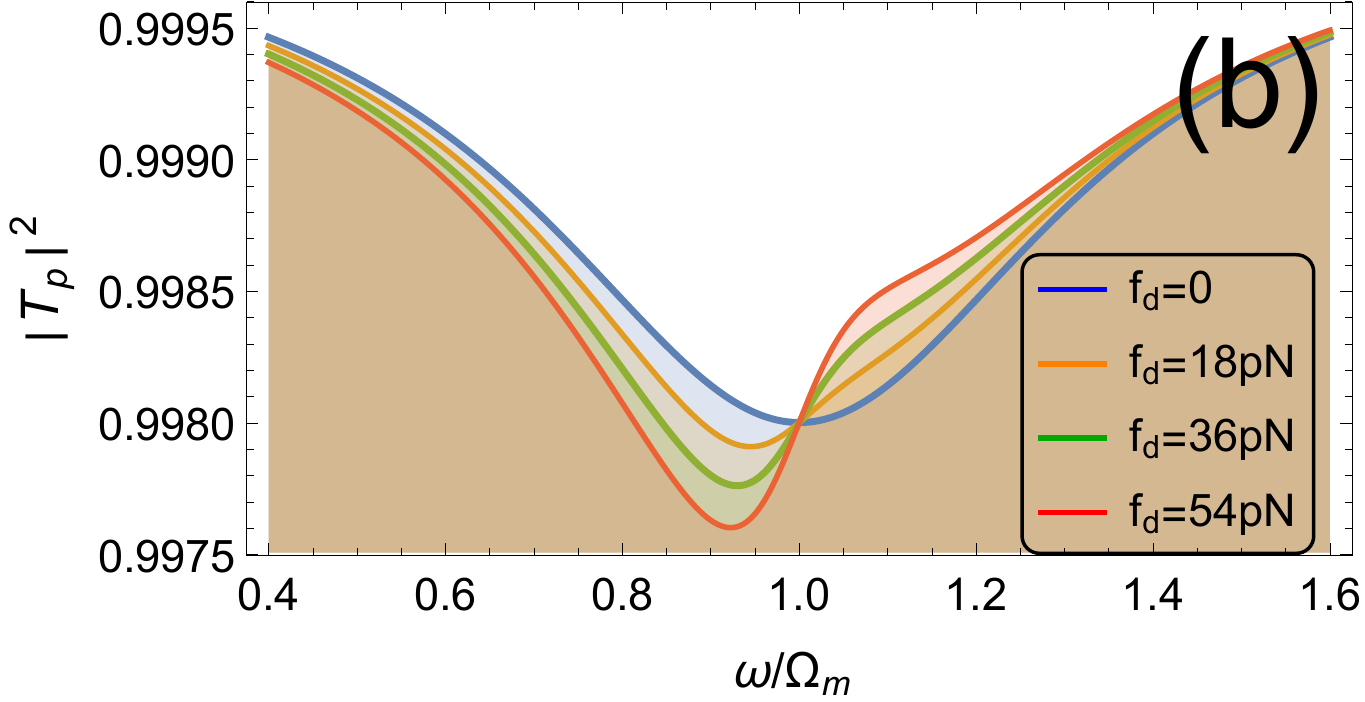}
\caption{\textbf{Micrutuble induced transparency.}~The probe transmission parameter $ |T_p|^2 $ versus normalized frequency $ \omega/\Omega_m $ and for different values of the force $ f_d $ and for (a) $ \phi=0 $ and (b) $ \phi=\pi/2 $. Here, we consider a micro-toroid silica cavity with refractive index $ n_c=1.44 $, wavelength $ \lambda_c=2\pi/\omega_c=1.55\mu$m, damping rate $ \kappa/2\pi\simeq \kappa_e/2\pi=20 $MHz, with negligible internal loss $ \kappa_i/2\pi=10 $kHz, circumference $ L_c=0.1$mm, radius $ R=1 \mu$m, and mode volume $V_c\sim 1.57\times 10^{-16} \mathrm{m}^3$. The decay length of the evanescent field is ~$ 1/k_{\perp}= 0.56\mu$m. We consider a doubly clamped MT with length $L= 1 \mu$m and the effective mass $ m\simeq 1.34 \times 10^{-19} $kg that is placed at a distance $  d=0.1\mu$m. The polarizability of the MT $ \alpha_{\parallel}=1.1\times 10^{-33} \,\mathrm{C m^2/V}$ which is correspond to the refractive index $ n_{MT}\simeq 1.6$. The MT fundamental vibrational frequency is $\Omega_m/2\pi= 20.68  $MHz with quality factor $ Q_m=\frac{\Omega_m}{\gamma_m}=3.44 $. The estimated MT-cavity field coupling rate is~$g_0\simeq 0.165 \mathrm{GHz/m}$. We also assume the cavity is coherently driven by an external laser at red-sideband $ \Delta=-\Omega_m $ whose input power $ P_l=500 \mu$W; corresponding to $ n_d\simeq 2.25\times 10^7 $ photons inside the cavity. The total probe photons inside the cavity is considered to be very small  $ n_{prob}=\langle \hat a^{\dagger}\hat a \rangle\simeq 2.37$. }
\label{figEIT}
\end{figure}

In Fig.~\ref{figEIT}(a) we plot the probe transmission parameter $ |T_p|^2 $~[Eq.~(\ref{transmision2})] versus normalized frequency $ \omega/\Omega_m $ for different values of the force $ f_d $ and for $ \phi=0 $ ~[Fig.~\ref{figEIT}(a)] and $ \phi=\pi/2 $~[Fig.~\ref{figEIT}(b)]. 

Here, we consider the experimentally feasible parameters: a micro-toroid silica cavity with refractive index $ n_c=1.44 $, wavelength $ \lambda_c=2\pi/\omega_c=1.55\mu$m, damping rate  $ \kappa/2\pi\simeq \kappa_e/2\pi=20 $MHz, with small internal loss $ \kappa_i/2\pi=10$kHz, circumference $ L_c=0.1$mm, radius $ R=1 \mu$m, and mode volume $V_c\sim 1.57\times 10^{-16} \mathrm{m}^3$~\cite{Vahala, Heebner, Kipp}. The cavity is placed inside a water bath with refractive index $ n_{en}=1.33 $  where the decay length of the evanescent field is ~$ 1/k_{\perp}= 0.56\mu$m. However, we consider a doubly clamped MT with length $L= 1 \mu$m and the effective mass $ m\simeq 1.347 \times 10^{-19} $kg that is placed at a distance $  d=0.1\mu$m from the cavity rim whose polarizability is $ \alpha_{\parallel}=1.1\times 10^{-33} \,\mathrm{C m^2/V}$~(corresponds to the refractive index $ n_{MT}\simeq 1.6$)~\cite{Cifra0}. We consider a MT with the fundamental vibrational frequency $\Omega_m/2\pi= 20.68  $MHz and damping rate $ \gamma_M/2\pi=6 $ MHz whose quality factor is $ Q_m=\frac{\Omega_m}{\gamma_m}=3.44 $. For these values the estimated MT-cavity field coupling rate is~$g_0\simeq 0.165 \mathrm{GHz/m}$. The cavity is coherently driven by an external laser at red-sideband $ \Delta=-\Omega_m $ whose input power $ P_l=500 \mu$W corresponds to $ n_d\simeq 2.25\times 10^7 $ photons inside the cavity. In contrast, the total probe photons inside the cavity is considered to be very small  $ n_{prob}=\langle \hat a^{\dagger}\hat a \rangle\simeq 2.37$. Note that in this paper we always limit our analysis to the case $ k_{\perp}(X_d+d) \ll 1 $ in which the MT displacement is small compared to the decay length~$ 1/k_{\perp} $ of the evanescent field. This condition imposes a constraint on the amplitude of the applied force $ f_d \ll \Big[(k_{\perp}^{-1}-d)m\gamma_m\Omega_m\Big]\simeq 0.3$nN. 

Figure~\ref{figEIT}(a) shows that in the absence of an external force~$ f_d=0 $, due to retardation effect of the cavity, a dip appears in the first MT's sideband $ \omega=\Omega_m $ while the MT vibration in the presence of the external force totally changes this behavior and leads to the appearance of a transparent window in the transmission of the probe field. In particular, we can see that the stronger driving force makes the dip of absorption
shallower and eventually the dip emerges in the peak, leading to optomechanically induced transparency of the probe field. 

In principle, the radiation pressure exerted by the probe field on the MT and the external driving
force results in the anti-Stokes scattering of light from the drive field, which produces two kinds of anti-Stokes fields induced by the probe field and the external force, respectively. Then the interference of these two anti-Stokes fields and the intracavity probe field leads to the transparency of the probe field and the appearance of the peak in the transmission of the probe. In other words, the external driving force coherently enhances the oscillation of the MT, leading to optomechanically induced transparency. We can realize the optomechanically induced transparency for a resonantly injected
probe in the MT-cavity system by appropriately adjusting the amplitude of the probe(or drive) and the external force. As seen from Eq.~(\ref{transmision2}), around MT frequency $\Omega_m  $, the magnitude of this transparency peak is given by $\frac{G_d}{\sqrt{\kappa_e}\mathcal{E}_p}  $.   

\begin{figure}[h]
\centering
\includegraphics[width=1\columnwidth]{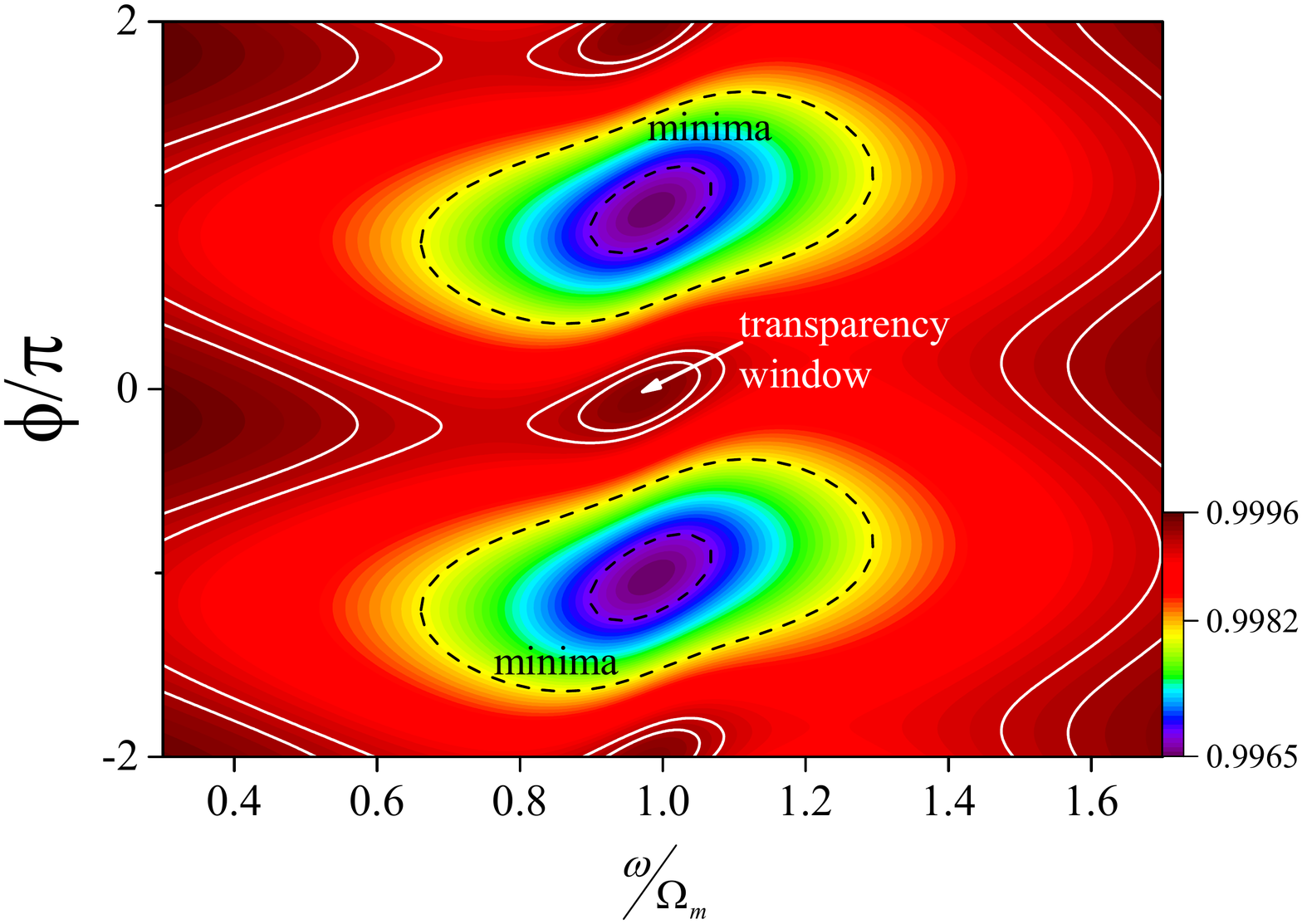}
\caption{\textbf{Transparency with respect to the relative phase. } The probe transmission parameter $ |T_p|^2 $ versus $ \omega/\Omega_m $ and relative phase $ \phi $ for $ f_d=60 $pN. Here, the other parameters are the same as Fig.~\ref{figEIT}.}
\label{figEIT3D}
\end{figure}
 
 However, the response of the system also tightly depends on the relative phase between the probe field and the external driving force~[see Fig.~\ref{figEIT}(b)]. In Fig.~\ref{figEIT3D} we plot the probe transmission parameter $ |T_p|^2 $ versus $ \omega/\Omega_m $ and relative phase $ \phi $. It is clear that the symmetric peaks and therefore the probe transparency appear around $ \phi=0 $. The center frequency and linewidth of the transparency peak, respectively, give the resonance frequency~$ \Omega_m $ and damping rate~$ \gamma_M $ of the MT while the height of the peak reveals information about $ g_0 $ and $ f_d $. On the other hand, the deep dips~(no transparency)  appear at $ \phi=\pm\pi $ while the asymmetric structures rises up when the relative phase~$\phi  $ approaches to $\pm \pi/2 $. In this case the probe transmission stands between the peaks of $ \phi=0 $ and the dips of $ \phi=\pm \pi $. 

\begin{figure}[h]
\centering
\includegraphics[width=0.99\columnwidth]{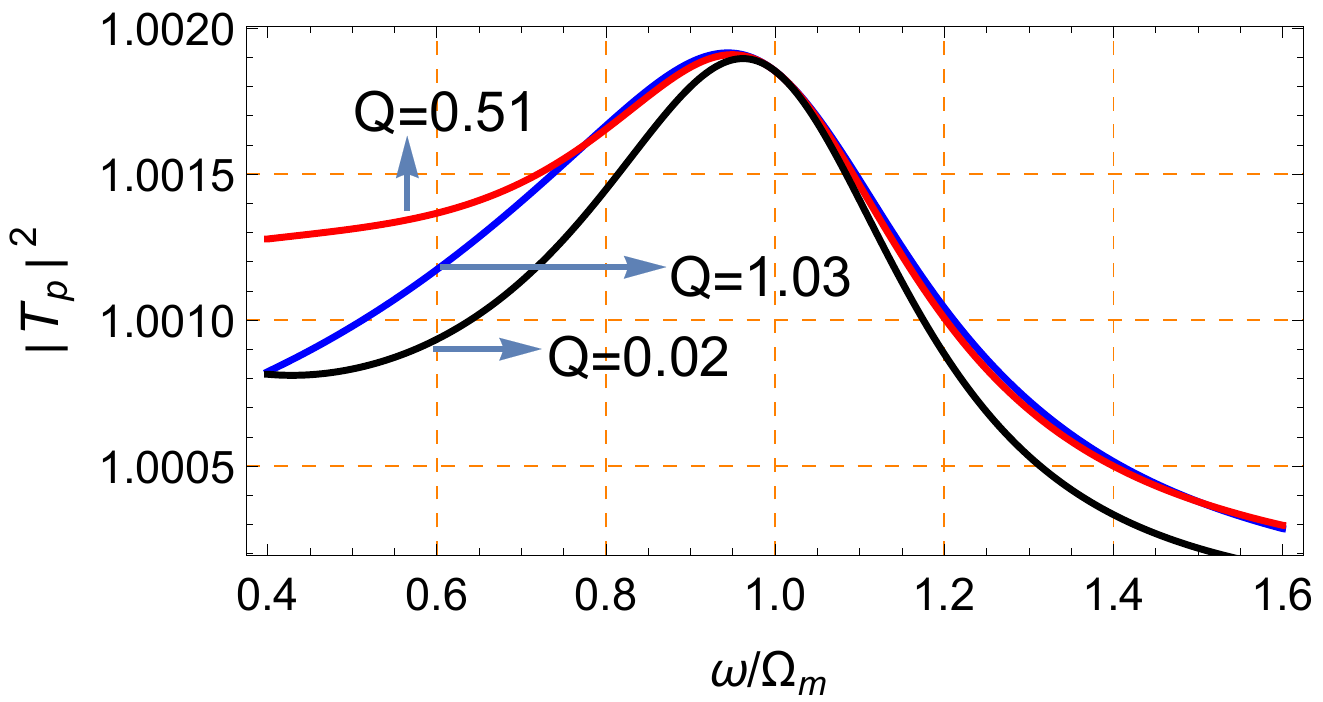}
\caption{\textbf{Effect of the quality-factor on transparency pattern}. The probe transmission parameter $ |T_p|^2 $ versus $ \omega/\Omega_m $ for different quality factors: $ Q=1.03\,(f_d=0.2\mathrm{nN}) $, $ Q=0.51\,(f_d=0.4\mathrm{nN}) $, and $ Q=0.02\,(f_d=10\mathrm{nN}) $ with $ \kappa_e/2\pi=10 $MHz and $ \kappa_i/2\pi=1 $kHz. Here, the other parameters are the same as Fig.~\ref{figEIT}.}
\label{quality}
\end{figure}

In Fig.~(\ref{quality}) we also study the effect of the different vibrational quality factor on the transparency pattern. Lower quality factor (higher damping rate) leads to suppression of the transparency peak. In order to keep the transparency peak one should vary the external driving force and increase it correspondingly. Note however, in Fig.~(\ref{quality}) we can only see the transparency peak because the MT's linewidth is much larger than the cavity linewidth i.e., $ \gamma_m\gg \kappa $. In this case the cavity is overwhelmed with the MT transparency window, therefore, the cavity response dip is totally covered by the transparency peak.  

 \blank
\textbf{Discussion.} ~~ In this paper, we have suggested new ways to control, manipulate and read out microtubule vibrations. 
The dielectric properties of microtubules provide a new possibility to control and manipulate the microtubule vibrations via the electrostatic gradient force from an inhomogeneous external electric field. This can be achieved by positioning tip electrodes close to the center of a doubly clamped microtubule. Information about the microtubule mechanical vibrations can be obtained by coupling the microtubule to an optical cavity. We have shown that the optomechanical coupling between the microtubule and the evanescent field of the optical cavity modifies the response of the cavity field in the presence of a strong pump laser on the red mechanical sideband, leading to the appearance of transparency peaks in the transmission of an optical probe field. Analyzing the center frequency and linewidth of the transparency peak, one can extract the resonance frequency and damping rate of the vibrational mode of the microtubule. By varying the parameters of the system, in particular the magnitude of the driving force, one can observe higher vibrational frequencies~(up to 1GHz). Having good knowledge about the broad spectrum of mechanical frequencies in microtubules can open new doors for cancer therapy via external physical signals instead of chemical drugs, in order to minimize damage to healthy cells while achieving high control of cancer cells.

\blank
\textbf{Acknowledgment.}
The work of S. B. has been supported by the European Union's Horizon 2020 research and innovation program under the Marie Sklodowska-Curie grant agreement No MSC-IF  707438 SUPEREOM. M.C. acknowledges funding from Czech Science Foundation, project no. P102/15-17102S, and participates in COST Action BM1309 and project no. SAV-15-22 between Czech and Slovak Academies of Sciences. J.A.T. and C.S. acknowledge research funding from NSERC (Canada). 
\blank
\\
\textbf{Author contributions.}
All authors contributed to developing the proposal and writing the manuscript.

\blank
\textbf{Methods}\\
\small{
\textbf{Microtubule equations of motion}.
In the absence of dissipation and other external forces, the dynamic behaviour of the flexural MT can be obtained by Lagrangian (\ref{lag}), which gives
\begin{equation}\label{equationmotion}
\mu \frac{\partial^2 y}{\partial t^2}+(\sigma^2 Y\,A) \frac{\partial^4 y}{\partial x^4}=0,
\end{equation} 
The eigenmodes of this equation are given by
\begin{eqnarray}\label{eigenmode}
\psi_n(x)&=&\frac{1}{C_n}\Big[\frac{\mathrm{sin}(\zeta_n x/L)-\mathrm{sinh}(\zeta_n x/L)}{\mathrm{sin}(\zeta_n)-\mathrm{sinh}(\zeta_n)}\nonumber\\
&&-\frac{\mathrm{cos}(\zeta_n x/L)-\mathrm{cosh}(\zeta_n x/L)}{\mathrm{cos}(\zeta_n)-\mathrm{cosh}(\zeta_n)}\Big],
\end{eqnarray}
where $ C_n $ represent the normalization constants and the eigenvalues $ \zeta_n $ satisfying the transcendental equation $ \mathrm{cos}(\zeta_n)\mathrm{cosh}(\zeta_n)=1 $, with solutions $ \zeta_n=4.73, 7.85,... $. As a result, one can expand the general solution of Eq. (\ref{equationmotion}) in terms of eigenmodes (\ref{eigenmode}), i.e.,
\begin{equation}\label{solution}
y(x,t)=\sum_n \psi_n(x)X_n(t). 
\end{equation}

Using the Lagrangian in Eq.~(\ref{lag}) along with Eq. (\ref{solution}) one obtains the Hamiltonian describing MT vibrations
\begin{equation}\label{hamiltonianH00}
H_0=\sum_n \Big(\frac{P_n^2}{2m_n}+\frac{1}{2}m_n\omega_n^2X_n^2\Big).
\end{equation}
\textbf{Estimating the MT-cavity field coupling rate.}
For simplicity, we consider only the motion and the elastic deformations of the MT taking place along the spatial direction $x$, orthogonal to its reflecting surface. The evanescent field of the optical cavity, with electric field $ \vec{E}(\vec r) $, generates an
interaction with the dipoles in the MT. In fact, the evanescent field of the cavity
exerts a radiation pressure force on the MT, which is proportional to its
intensity and its phase is simultaneously shifted by the MT displacement from the equilibrium position. In the limit of small MT displacements, the
coupling part of the Hamiltonian is given by~\cite{hartman}
\begin{equation}\label{optomechanicalHamiltonian}
\hat H_{MC}=-\frac{1}{2}\int_{V_{MT}}\hat{\vec{P}}(\vec r).\hat{\vec{E}}(\vec r)\,dV,
\end{equation}
where $\hat{\vec{P}}(\vec r)=\vec{\alpha}.\hat{\vec{E}}(\vec r) $ is the polarization vector with $ \vec{\alpha} $ being the screened polarizability tensor of the MT. Here, the integration is applied over the MT's volume $ V_{MT} $. 

In order to find the optomechanical coupling rate between the MT and the cavity field, we represent the electric field in its quantized form
\begin{equation}\label{Eletricfield}
\hat{\vec{E}}(\vec r)=\sqrt{\frac{\hbar \omega_c}{2\epsilon_0}}(\hat a+\hat a^{\dagger})\psi_{\phi}(\vec r)\hat \phi,
\end{equation}
where $ \epsilon_0 $ is the vacuum permittivity and $ \vec{\psi}(\vec r)= \psi_{r}(\vec r)\hat r+\psi_{\theta}(\vec r)\hat \theta+\psi_{\phi}(\vec r)\hat \phi $ is the normalized eigenmode of the field, satisfying
\begin{equation}
\int \frac{\epsilon(\vec r)}{\epsilon_0}|\vec \psi(\vec r)|^2 \,dV=1, 
\end{equation}
For a toroidal microcavity of radius $ R $, the normalized eigenmode of the field outside the cavity ($ r>R $) in $ \hat \phi$ direction is given by \cite{Heebner, Kipp}
\begin{equation}
\psi_{\phi}(\vec r)\simeq \frac{-0.42\lambda_c}{R\sqrt{n_c^2-n_{en}^2}n_c\sqrt{V_c}}\sqrt{\frac{R}{r}}e^{-k_{\perp}(r-R)},
\end{equation}
where $ \lambda_c=2\pi c/\omega_c $ and $ n_c $ are the wavelength and refractive index of the microcavity, respectively, and  $k_{\perp}^{-1}=1/\sqrt{n_c^2-n_{en}^2}k $ is the decay length of the evanescent field, with $ k=2\pi/\lambda_c $ being the wavenumber of the field. $n_{en}  $ is the refractive index of the cavity/MT environment and $ r $ is the radial distance from the center of the cavity. We employ the following definition for the mode volume
\begin{equation}\label{modevol}
V_c=\int \frac{\epsilon(\vec r)}{n_c^2\epsilon_0}\frac{|\vec{E}(\vec r)|^2}{|\vec{E}_{max}|^2} \,dV,
\end{equation}
where $ \vec{E}_{max} $ stands for the maximum of the electric field and $ L_c $ is the circumference of the cavity. 

We assume that the MT displacement~$X$ is small compared to the decay length~$ 1/k_{\perp} $ of the evanescent field. Therefore, we consider the electric field to be constant inside the MT volume $ V_{MT} $ and we can linearize the optomechanical coupling $ \vec E. \vec{\alpha}. \vec E $ around the equilibrium position of the MT. Substituting Eqs.~(\ref{Eletricfield})-(\ref{modevol}) into Eq.~(\ref{optomechanicalHamiltonian}) and comparing the result with Hamiltonian~(\ref{Hamtotal}) gives the MT-cavity field a coupling rate which is approximated as
\begin{equation}\label{couplingg00}
g_0\simeq\Big(\frac{\omega_c \alpha_{||}k_{\perp}\zeta^2 }{n_c^2 \epsilon_0 V_c}\mathrm{e}^{-2k_{\perp}d}\Big)A_{c}
\end{equation}
where $ \zeta=\frac{0.42 \lambda_c}{R\sqrt{n_c^2-n_{en}^2}} $ and $ d $ is the distance between the MT center and the cavity rim. The correction term $ A_c\simeq 0.17[k_{\perp}(d+R)]^{-1/2} $ accounts for the misalignment and mispositioning of the MT with respect to the cavity rim~\cite{hartman}. In the derivation of Eq.~(\ref{couplingg00}), we have neglected the contribution of the perpendicular polarizability as we assume the parallel polarizability of the MT provides the main contribution to the optomechanical coupling. 

\blank
\textbf{Response of the system.}
To solve the equations~(\ref{meanEqution}) we take account of the first-order sidebands and ignore the higher order sidebands by considering the following ansatz
\begin{subequations}\label{Solutions}
\begin{eqnarray}
\langle \hat a\rangle &=& \sqrt{n_d}+\alpha_1^- e^{-i\delta t}+\alpha_1^+e^{i\delta t}+\alpha_2^- e^{-i\omega t}+\alpha_2^+e^{i\omega t},\nonumber\\
\\
\langle \hat X\rangle &=& x_0+x_1e^{-i\delta t}+x_1^*e^{i\delta t}+x_2e^{-i\omega t}+x_2^*e^{i\omega t},\nonumber\\
\end{eqnarray}
\end{subequations}
By substituting the ansatz Eq.~(\ref{Solutions}) into Eqs.~(\ref{meanEqution}) and retaining the first-order terms, we can find all unknown parameters in Eq.~(\ref{Solutions}). 

The output field of the cavity, in the laboratory frame (unrotated frame), can be obtained by using the input-output relation
\begin{eqnarray}\label{output}
\langle \hat a_{out}\rangle&=&\big(\mathcal{E}_le^{-i\omega_l t}+\mathcal{E}_p e^{-i\omega_p t-i\phi_p}\big)-\sqrt{\kappa_e} \langle \hat a \rangle \\
&=&(\mathcal{E}_l-\sqrt{\kappa_e}\sqrt{n_d})e^{-i\omega_l t}+(\mathcal{E}_p\,e^{-i\phi_p}-\sqrt{\kappa_e}\alpha_1^-)e^{-i\omega_p t}\nonumber\\
&-&\sqrt{\kappa_e}\Big(\alpha_1^+e^{-i(2\omega_l-\omega_p)t}+\alpha_2^-e^{-i(\omega_l+\omega)t}\nonumber\\
&+&\alpha_2^+e^{-i(\omega_l-\omega)t}\Big),\nonumber
\end{eqnarray}
where 
\begin{eqnarray}
 x_0&=&-\frac{\hbar g_0 n_d+\bar{F}_0}{m\Omega_m^2},\,\alpha_1^-= \frac{\Big(1+\frac{i\hbar g_0^2n_d\chi_m(\omega)}{i(\Delta-\omega)+\frac{\kappa}{2}}\Big)\sqrt{\kappa_e}\mathcal{E}_pe^{-i\phi_p}}{\frac{2i\hbar g_0^2n_d\chi(\omega)\Delta}{i(\Delta-\omega)+\frac{\kappa}{2}}+\frac{\kappa}{2}-i(\Delta+\omega)},\nonumber\\
\alpha_2^- &=& \frac{1}{2}\frac{-ig_0\sqrt{n_d}\chi_m(\omega)\,f_d\,e^{-i\phi_d}}{\frac{2i\hbar g_0^2n_d\chi_m(\omega)\Delta}{i(\Delta-\omega)+\frac{\kappa}{2}}+\frac{\kappa}{2}-i(\Delta+\omega)}.\nonumber
\end{eqnarray}
Here, $ n_d = \frac{4\kappa_e^2\mathcal{E}_l^2}{\kappa^2+4\Delta^2}$ shows the total number of photons inside the cavity with $ \Delta=\Delta_0-g_0x_0 $ being the effective optical detuning. Note that, in Eq.~(\ref{output}), the terms related to $ \alpha_1^{\pm} $ introduce two sidebands induced only by the MT and driving pump field while the terms associated with $ \alpha_2^{\pm} $ determine the extra sidebands induced by the external driving force acting on the MT. 

In this paper we are interested in the transmission of the probe field which is the ratio of the probe field returned from the system divided by the sent probe field and it is given by~\cite{Weis, Safavi}
\begin{eqnarray}\label{transmision}
T_p=\frac{\langle \hat a_{out}\rangle}{\mathcal{E}_p e^{-i(\omega_pt+\phi_p)}}.
\end{eqnarray}
By plugging Eq.~(\ref{output}) into Eq.~(\ref{transmision}) and neglecting the fast rotating terms at $ \Omega_M $ and for $ \omega_p=\omega_c $ and $ \omega=\Omega_m $, we obtain
\begin{eqnarray}\label{transmision23}
T_p(\omega)&=&1-\frac{\sqrt{\kappa_e}}{\mathcal{E}_pe^{-i\phi_p}}\Big(\alpha_1^-+\alpha_2^-\Big).
\end{eqnarray}

\end{document}